\documentclass[nofootinbib,prx,twocolumn,amsmath,amssymb,aps,]{revtex4-1}
\usepackage{graphicx}
\usepackage{hyperref}

\usepackage{float}
\hypersetup{
     colorlinks   = true,
     linkcolor    = blue,
     citecolor    = blue
}
\usepackage[all]{hypcap}

\begin{document}

\title{Almost linear Haldane pseudopotentials and emergent conformal block wave-functions in a Landau level}
\author{Yahya Alavirad$^1$}
\affiliation{
$^1$Department of Physics, Condensed Matter Theory Center and the Joint Quantum Institute, University of Maryland, College Park, MD 20742, USA}

\date{\today}

\begin{abstract}
We consider a two dimensional (2D) model of particles interacting in a Landau level. We work in a finite disk geometry and take the particles to interact with a linearly decreasing two-body Haldane pseudo-potential. We show that the ground state subspace of this model is spanned by the wave-functions that can be written as polynomial conformal blocks (of an arbitrary conformal field theory) consistent with the filling fraction (scaling dimension). To remove degeneracies, we then add a quadratic perturbation to the Hamiltonian and show that; 1. Conformal blocks constructed using the Moore-Read construction (e.g. Laughlin, Pfaffian, and Read-Rezayi states) remain \textit{exact eigenstates} of this model in the thermodynamic limit and 2. By tuning an externally imposed single-body $-L_z^2$ potential we can enforce Moore-Read conformal blocks to become \textit{exact ground states} of this model in the thermodynamic limit. We cannot rule out the possibility of residual degeneracies in this limit. This model has no filling dependence and is comprised only from two-body long-range interactions and external single-body potentials. Our results provide insight into how conformal block wave-functions can emerge in a Landau level.
\end{abstract}

\maketitle

\section{Introduction}
Non-interacting two dimensional electrons in a uniform magnetic field form flat Landau levels (LLs)\cite{landau}. In strong magnetic fields where typical interaction energy scale becomes negligible compared to the LL spacing, the full (interacting) problem can be projected into a single LL. When partially filled, interactions in a LL can give rise to the novel phenomena of fractional quantum Hall (FQH) effect. 
 
 A significant part of the theoretical understanding of the FQH effect relies on model wave-functions. These wave-functions accurately capture essential features of the FQH effect. They are not exact ground state of any realistic physical system. Yet, exact diagonalization for small systems shows that at least some of them (e.g. the Laughlin $1/3$ state), have remarkably large overlaps with the true ground state of electrons interacting with the Coulomb interaction. Following the seminal work of Laughlin\cite{laughlin}, several classes of important model wave-functions have been proposed. Examples of which include; Haldane-Halperin hierarchy\cite{Haldane83,Halperin84}, Jain's composite fermions\cite{jain7} and conformal block wave functions introduced by Moore and Read\cite{mooreread,rr,rmp} (this includes the Laughlin, Pfaffian, Read-Rezayi and the Jack polynomial states\cite{jack1,jack2}). Moore and Read's conformal block wave-functions will be the focus of this work. They are constructed as certain conformal blocks of chiral $U(1)\otimes \Psi$ conformal field theories (CFTs). Here the $U(1)$ part is called the ``charge" sector and the $\Psi$ part is called the ``statistics" sector. Different choices of the statistics sector correspond to different trial states (e.g. if $\Psi$ is the free fermion CFT we get the Pfaffian state). While these wave-functions are exact eigenstates of certain Haldane pseudopotential models - these models are filling dependent and include higher body terms.
 
 In this paper, we work in two-dimensional disk geometry of radius $R$. We consider a simple model of particles in a LL interacting with linearly decreasing (two-body) Haldane pseudopotentials and show that, quite remarkably, the ground state subspace of this model is given by conformal blocks (correlators of identical primary fields) of chiral conformal field theories (CFTs) invariant under global conformal transformations $SL(2,\mathbb{C})$. The only restrictions on these conformal block wave functions is that they need to 1. Be consistent with the filling fraction (primary fields need to have a certain scaling dimension). 2. Be polynomial (single-valued) and 3. Respect the proper statistics (e.g. Fermi or Bose). We then add a (physically motivated) small quadratic perturbation to the pseudopotential to remove artificial degeneracies. The effect of this quadratic term can be equivalently seen as choosing a certain CFT from the space of all CFTs where a consistent conformal block exists. We show that the resulting model has conformal blocks constructed using Moore and Read's construction (e.g. Laughlin, Pfaffian, Read-Rezayi states) as it's \textit{exact eigenstates} in the thermodynamic limit. We further demonstrate that by tuning an external single-body potential $-L_z^2$ one can further tune the system to ensure that the Moore-Read conformal block wave-functions become \textit{exact ground states} in the thermodynamic limit (though we cannot rule out the possibility of residual degeneracies). As opposed to other pseudopotential models, our model is entirely comprised of (long-range) two-body interactions and single-body potentials. Moreover, it has no filling dependence and can, therefore, be used for all filling fractions simultaneously, potentially, allowing one to study FQH transitions analytically. Our results provide further insight into the emergence of conformal block wave-functions in a LL and can serve as a starting point to perturbatively study more general pseudo-potential models using connections with CFT. 
 
 The rest of this paper is organized as follows: Section.\ref{efh1} introduces the model and the formalism we use. In Section.\ref{psd1} we diagonalize the linear pseudopotential model and demonstrate its relationship with conformal blocks of chiral CFTs. In Section.\ref{psd2} we add the quadratic perturbation and discuss it's relation with Moore-Read's conformal block wave-functions. We end with a brief discussion and conclusion in Section.\ref{disc}. The Supplementary Material.\ref{num} includes sample numerical results from exact diagonalization on small systems.
 
\section{Effective Hamiltonian in a Single Landau Level}\label{efh1}

 \begin{figure}
\centering
\includegraphics[width=\columnwidth,keepaspectratio]{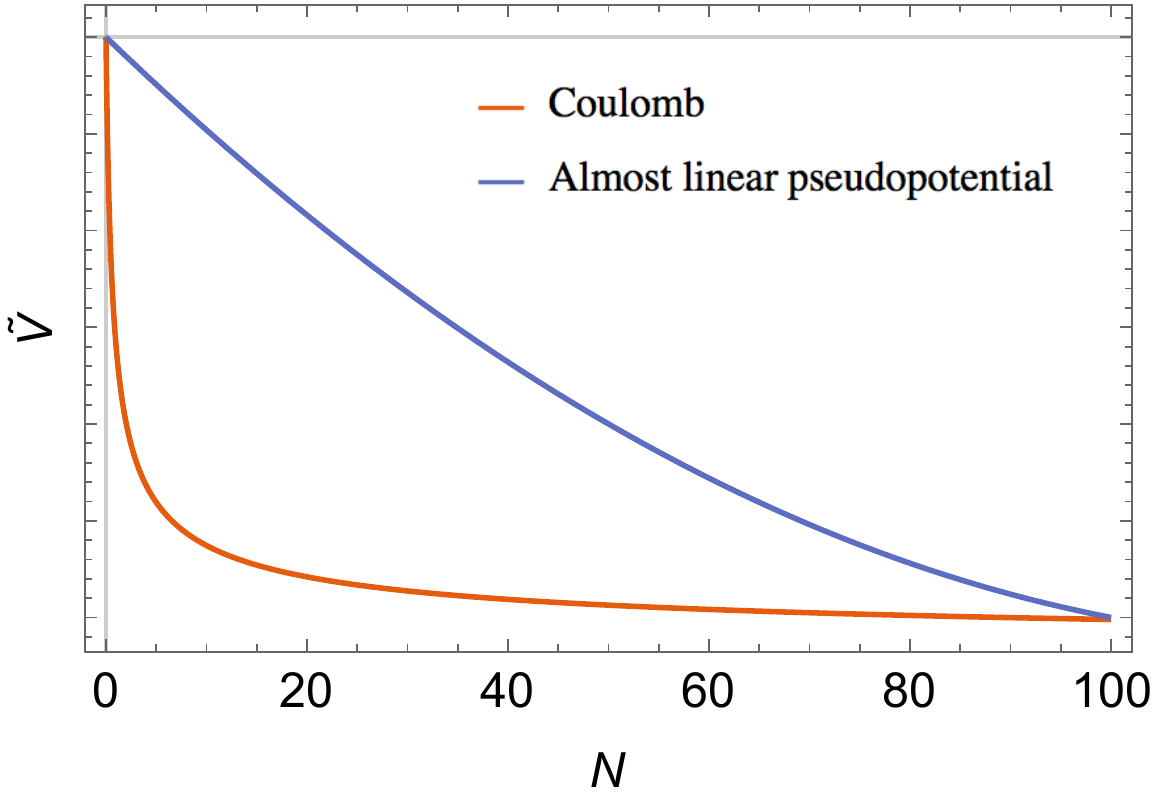} 
\caption{Haldane pseudopotentials for Coulomb interaction projected to the LLL (red), and the almost linear pseudopotential model (blue). \label{fig:1.pdf}}
\end{figure}
We work in a two-dimensional (2D) rotationally invariant disk geometry. The Hamiltonian describing $N_e$ electrons interacting in a magnetic field is,
\begin{align}
H = \frac{1}{2m_e} \sum_{j=1}^{N_e} \mathbf{\pi}_j^2 + \sum_{\mathbf{q}}V(|\mathbf{q}|)\sum_{j<k} e^{i \mathbf{q}.(\mathbf{r}_j-\mathbf{r}_k)},
\end{align}
where $N_e$ is the number of particles, $m_e$ is the electron mass and $\mathbf{\pi}=\mathbf{P}-e \mathbf{A}$ is the canonical momentum. In a uniform magnetic field $B$ we get $[r_j^a,\pi_k^b]=i \hbar \delta_{jk} \delta_{ab}$ and $[\pi_j^a,\pi_k^b]=i \delta_{jk}\epsilon_{ab} \frac{\hbar^2}{l_B^2}$. Here $l_{B}=\sqrt{\frac{\hbar}{eB}}$ is the magnetic length. Motivated by classical analogy one can define the ``guiding center coordinates", $R_j^x=r_j^x + \pi^y_j \frac{l_B^2}{\hbar},  \quad R_j^y=r_j^y - \pi^x_j \frac{l_B^2}{\hbar}$. These coordinates define a non commutative geometry $[R_j^a,R_k^b]=-i  l_B^2 \epsilon_{ab} \delta_{jk}$. Guiding center coordinates commute with the kinetic term in the Hamiltonian $[R_j^a,\pi_k^b]=0$. At zero temperature and in the large magnetic field limit $ V \ll \frac{eB\hbar}{m}$, the entire problem can be mapped into the lowest partially filled landau level, in this limit only the guiding center degrees of freedom survive. The effective single LL Hamiltonian is then given by,
\begin{align}
H_e= \sum_{\mathbf{q}} V_e({|\mathbf{q}|}) \sum_{j<k} e^{i \mathbf{q}.(\mathbf{R}_j-\mathbf{R}_k)}.\label{h0}
\end{align}
Here $V_e$ is the projected form of interaction into a fixed LL. From the guiding center coordinates we can define ladder operators $a_j=\frac{1}{\sqrt{2}l_B} (R_j^x-i R_j^y)$, that satisfy the usual $[a_j,a_k^\dagger]=\delta_{j,k}$ commutation relations. To develop intuition, notice that by assuming rotational invariance and within the symmetric gauge, the number operator $a_j^\dagger a_j$ counts the $L_z$ component of the total angular momentum of the $j'th$ particle $L_j^z$ (shifted by a LL dependent constant). Note that (assuming rotational symmetry) the two-body Hamiltonian preserves the ``relative angular momentum". Using these ladder operators and assuming rotational symmetry (to help perform the azimuthal part of the $\mathbf{q}$ integral), we can rewrite the Hamiltonian in Eq.\eqref{h0} as,
\begin{align}
H_e= \sum_{j<k} U(N_{jk}),\label{h1}
\end{align}
with 
\begin{align}
N_{jk}= \frac{1}{2} (a^\dagger_j-a^\dagger_k)(a_j-a_k),
\end{align} 
defining the (integer valued) ``relative angular momentum" number operator. The function $U(N)$ is the well-known Haldane pseudo-potential (i.e. energy cost of having a pair of particles at relative angular momentum $N$). The form of $U(N)$ for Coulomb interaction projected to the lowest LL is shown in Fig.~\ref{fig:1.pdf} (shifted by an inconsequential constant).

Note that the total Hilbert space is the space of $N_e$ decoupled Harmonic oscillators. This Hilbert space is spanned by polynomials of creation operators applied to the vacuum state (the state that is annihilated by all annihilation operators $a_i$). Whereas, the physical Hilbert space, associated with e.g. spin-less fermions (anti-symmetric wave-functions) corresponds to a subset of that space, spanned by anti-symmetric polynomials of creation operators applied to the ground state. In this work, we do not restrict to symmetric/anti-symmetric wave-functions, however, all calculations in this paper can be easily generalized to the case of spin-less bosons or fermions.

 We now turn to discussing the effect of having a physical boundary. We model a physical disc of radius $R$, by adding an large step function barrier potential that prohibits particles from having an angular momentum larger than $M=\frac{R^2}{2l_B^2}$ (a single particle wave function with angular moment $m$ is peaked at $R=\sqrt{2M} l_B$). To avoid complications of LL mixing we assume that the height of the potential barrier $W$ satisfies $U \ll W \ll \frac{eB\hbar}{m_e}$. The effective Hamiltonian then becomes,
 \begin{align}
H_{eff}(R=\sqrt{2M} l_B)= \sum_{j<k} U(N_{jk})+ W \sum_{j} \Theta(a^\dagger_j a_j-M). \label{h2}
\end{align}
Here $\Theta$ is the unit step function. The second term effectively enforces each electron to have an angular momentum smaller than $M$. Without loss of generality, from here on we take $U$ and $W$ to be dimensionless, i.e. we make the Hamiltonian dimensionless.

\section{Almost Linear Pseudopotentials}

 The form of $U(N)$ for Coulomb interaction is too complicated to study analytically. However, studying simple, yet,  ``physically reasonable" pseudopotentials can still provide useful insight. 
 
 The simplest form of an analytically tractable pseudo-potential is a straight line $U(N)=-N$. The negative sign of the slope makes the interaction repulsive. Using this form we get,
 \begin{align}
H_{L}(R=\sqrt{2m} l_B)= -\frac{2}{N_e} \sum_{j<k} N_{jk}+ W \sum_{j} \Theta(a^\dagger_j a_j-M). \label{hl}
\end{align}
 The factor of $1/N_e$ is added to make the total energy of order $\mathcal{O}(N_e^2)$. This form is exactly solvable because the first term is bilinear in creation and annihilation operators $a^\dagger,a$. All non-trivialities that arise are a consequence of the constraint enforced by the second term. We remark that this problem is still a strongly interacting problem, and that exact solvability is a remarkable consequence of using guiding center coordinates and working in first quantization. 
 
 Since $U(N)=-N$ is in principle not bounded from below, one might worry about the stability of this Hamiltonian. However, note that presence of a physical boundary (Second term in Eq.\eqref{h2}), imposes an effective external upper bound on $N$. One way to explicitly ensure stability without changing the low energy physics is to deform $U(N)$ to a constant for large values $N>\frac{R^2}{2l_B^2}$(corresponding to electrons outside the droplet). Keeping this in mind, we shall ignore questions of stability from now on. 

 In the next section, we will find the low energy spectrum of $H_{L}$ in Eq.\eqref{hl} and show that the highly symmetric nature of it will result in artificial degeneracies in the spectrum. To remove these degeneracies we add a small quadratic perturbation to $U(N)$ to make it look like,
  \begin{align}\label{lin}
U_{Q}(N)=  - \frac{2}{N_e}N + \frac{4\alpha}{N_e^2}N^2.
\end{align}
Where $\alpha\ll1$ is a small dimensionless parameter. We shall treat the second term as a small perturbation and study the resulting ``almost linear" pseudopotentials using perturbation theory. A schematic figure of an almost linear pseudopotential, as well as its comparison with the Coulomb interaction in lowest LL, is given in Fig.~\ref{fig:1.pdf}.

Here, we emphasize that the method used in this work is quite general and can in principle be applied to any ``arbitrary" form of perturbation added to the linear term.

\subsection{Linear Pseudopotentials}\label{psd1}
We now proceed to diagonalize the linear pseudopotential model $H_{L}$ in Eq.\eqref{hl}. We are interested in the low energy part of the spectrum $E-E_0\ll W$. To this end, we focus on states that satisfy,
\begin{align}\label{cnst}
\Theta(a^\dagger_j a_j-m) | \psi \rangle = 0.
\end{align} 
 This is equivalent to constraining the electrons to reside entirely within the physical disk.

We start with the linear pseudopotential without the boundary term,
\begin{align}
 -\frac{2}{N_e} \sum_{j<k} N_{jk} = -\sum_{i=1}^{N_e} a^\dagger_i a_i + (\frac{\sum_{i=1}^{N_e}a^\dagger_i}{\sqrt{N_e}})(\frac{\sum_{i=1}^{N_e}a_i}{\sqrt{N_e}}).
 \end{align}
 In the equation above, energy is independent of the center of mass angular momentum $ (\frac{\sum_{i=1}^{N_e}a^\dagger_i}{\sqrt{N_e}})(\frac{\sum_{i=1}^{N_e}a_i}{\sqrt{N_e}})$. Whereas, all linear combinations of creation operators that commute with the center of mass angular momentum (e.g. $a^\dagger_1- a^\dagger_2$) decrease the energy of an eigenstate by $1$. A useful over-complete set of such operators is given by $a^\dagger_i-a^\dagger_j$. Using these operators, \textit{all} energy eigenstates can be written as a \textit{non-unique} homogeneous polynomial of degree $n$, $P_\text{h}(a^\dagger_i-a^\dagger_j)$ multiplied by a polynomial of center of master creation operator $P_\text{com}(\sum_i a^\dagger)$ applied to the vacuum state,
 \begin{align} \label{wvf}
 P_\text{com}(\sum_i a^\dagger_i) P_\text{h}(a^\dagger_i-a^\dagger_j) |0\rangle, 
 \end{align}
 the energy of this state is then given by $-n$ ($n$ is  the degree of the homogenous polynomial $P_h$).
 
  Without the boundary term the ground state energy of this Hamiltonian is not bounded from below. To obtain the ground state of the full Hamiltonian (with the boundary term), we can start with the vacuum state $|0\rangle$ and excite as many modes as possible without leaving the physical boundary, i.e. constraining the wave function $|\psi\rangle$ to satisfy $W \sum_{j} \Theta(a^\dagger_j a_j-M) |\psi\rangle =0$. Exciting the center of mass angular momentum mode effectively decreases the system size from $M\rightarrow M-1$ without changing the energy. Therefore, in the ground state, the center of mass angular momentum mode should not be excited $\sum_{i=1}^{N_e}a_i |\psi_0\rangle=0$. This is equivalent to setting $P_\text{com}(\sum_i a^\dagger_i)=1$ in Eq.\eqref{wvf}.
 
  We introduce a diagrammatic representation of the energy eigenstates of the form $P_\text{h}(a^\dagger_i-a^\dagger_j) |0\rangle$, (setting $P_\text{com}(\sum_i a^\dagger_i)=1$; as noted earlier ground state can be written in this form). We construct diagrams as follows; starting with the vacuum state, we represent the states $\prod_{i,j} (a^\dagger_i-a^\dagger_j)^{n_{ij}} |0\rangle$ by drawing $n_{ij}$ lines between the two points $i,j$ (to avoid a ``sign" ambiguity, we always take $i<j$). For example,
 \begin{align}
 (a^\dagger_i-a^\dagger_j) |0\rangle =\vcenter{\hbox{\includegraphics[width=0.3\columnwidth,keepaspectratio]{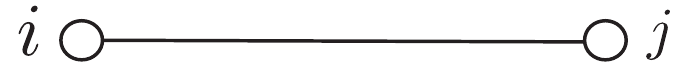}}}.
 \end{align}
 More complicated states can be represented similarly. For example,
  \begin{align}
 (a^\dagger_i-a^\dagger_j)^2 (a^\dagger_i-a^\dagger_k) (a^\dagger_j-a^\dagger_k)  |0\rangle =\vcenter{\hbox{\includegraphics[width=0.3\columnwidth,keepaspectratio]{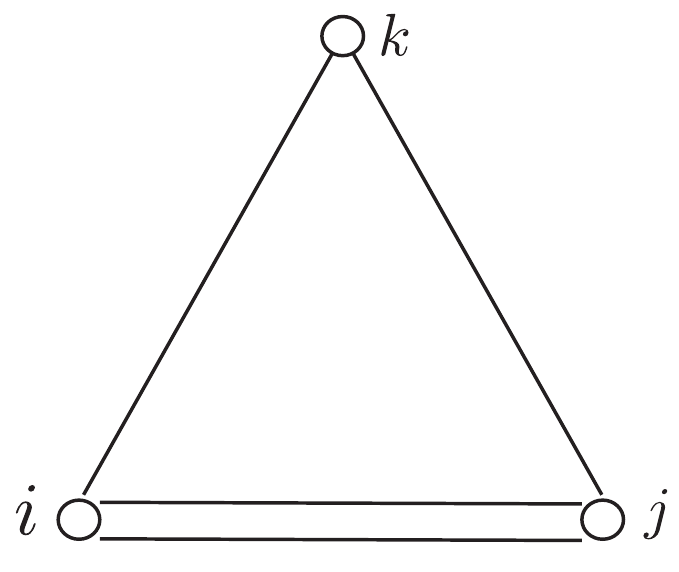}}}.
 \end{align}
 or,
  \begin{align}
 (a^\dagger_i-a^\dagger_k&) (a^\dagger_j-a^\dagger_k)  +  (a^\dagger_i-a^\dagger_j) (a^\dagger_i-a^\dagger_k)  |0\rangle =\\ \newline \nonumber
 &\vcenter{\hbox{\includegraphics[width=0.63\columnwidth,keepaspectratio]{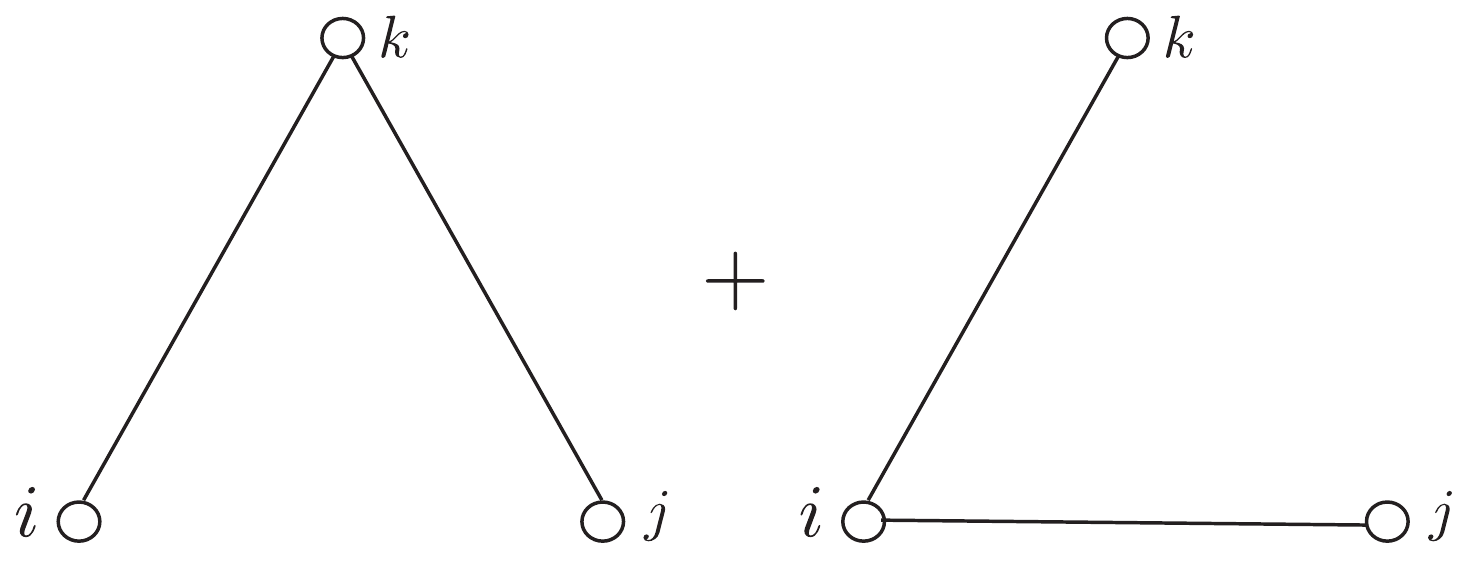}}}.
 \end{align}
 The advantage of using this diagrammatic form is that it makes it easy to enforce the boundary constraint of Eq.\eqref{cnst}. To see this, note that in these diagrams the maximum angular momentum of the $j'th$ particle is given by the number of lines attached to the vertex $j$. Therefore, enforcing the constraint of Eq.\eqref{cnst} is equivalent to working with diagrams that have equal or less than $M$ lines attached to each vertex. The strategy for finding the ground state is now simple; For $N_e$ particles at a system size $M$, we start by $N_e$ vertices representing each particle and then add as many lines between vertices as possible, without attaching more than $M$ lines to each vertex. As we'll show below, such states are given by polynomial conformal blocks of CFTs.
 
 We start by explicitly constructing the ground state in some simple cases; 
 
 1. For two particles, the \text{unique} ground state is trivially given by $(a^\dagger_1-a^\dagger_2)^M|0\rangle$ (for spin-less fermions the exponent $M$ changes to closest odd integer $M_o\le M$). For example for two particles at $1/3$ filling the ground is given by,
 \begin{align}
 \vcenter{\hbox{\includegraphics[width=0.4\columnwidth,keepaspectratio]{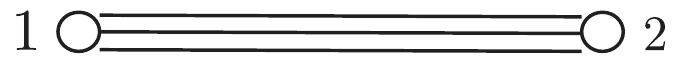}}}~~\text{for}~~N_e=2,~M=3(N_e-1)=3.
 \end{align}
 
 2. For three particles, at odd denominator fillings (even denominator for bosons), e.g. $1/3$. The \text{unique} ground state is given, \text{exactly}, by the famous Laughlin state, e.g. $\prod_{i<j}(a^\dagger_i-a^\dagger_j)^3|0\rangle$ for $1/3$ filling,
  \begin{align}
 \vcenter{\hbox{\includegraphics[width=0.4\columnwidth,keepaspectratio]{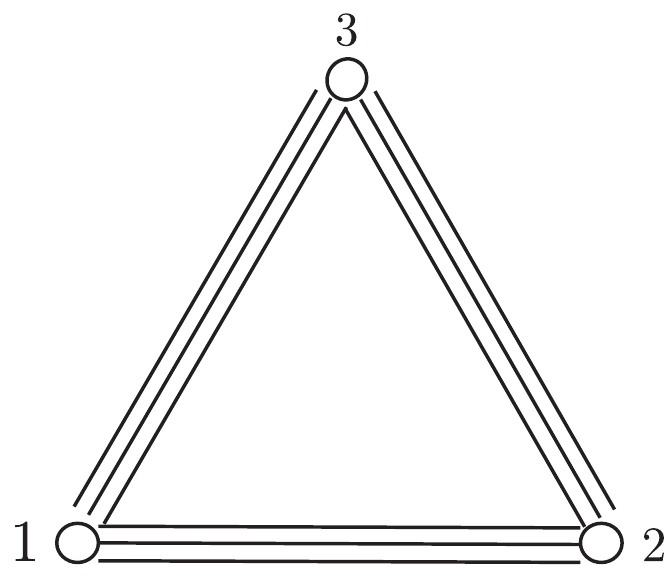}}}~~\text{for}~~N_e=3,~M=3(N_e-1)=6.
 \end{align}
 Note that this interesting result does not hold for generic pseudopotentials, e.g. the Coulomb interaction. 
 
 3. Four particles, at odd denominator fillings (even denominator for bosons), e.g. $1/3$. In this case the situation is more complicated as the ground state in no longer unique (it is degenerate). To see this, consider the $1/3$ filling case; Again, the Laughlin state corresponds to a ground state, 
   \begin{align}
 \vcenter{\hbox{\includegraphics[width=0.3\columnwidth,keepaspectratio]{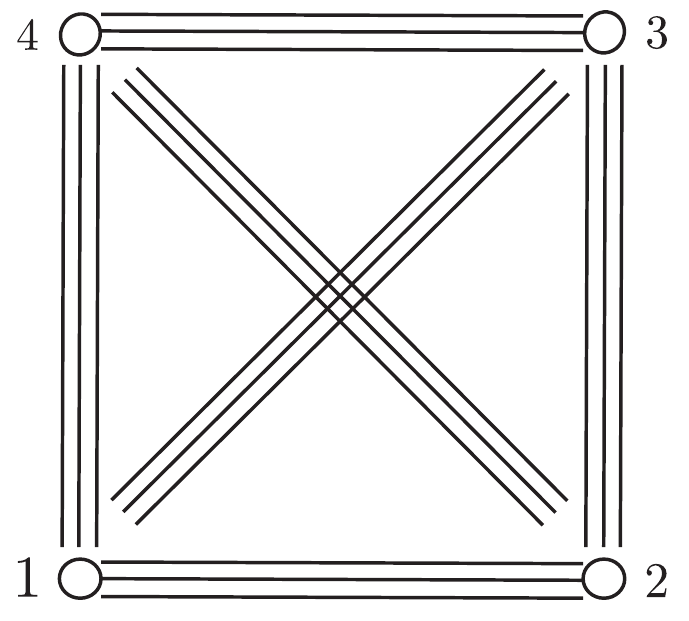}}}~~\text{for}~~N_e=4,~M=3(N_e-1)=9.
 \end{align}
However, in this case there are additional states (diagrams) with the same energy, for example,
   \begin{align}
 &\vcenter{\hbox{\includegraphics[width=\columnwidth,keepaspectratio]{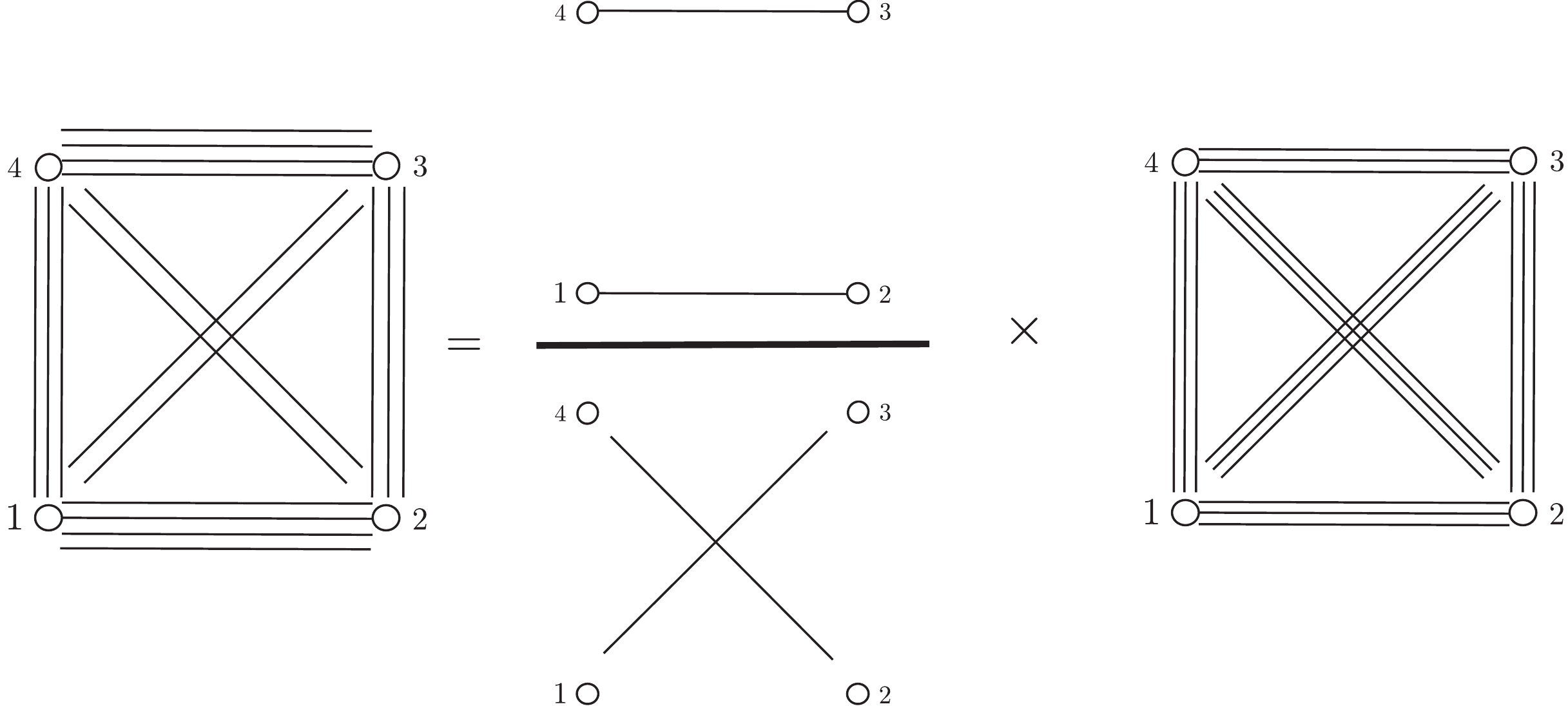}}}\\ \newline \nonumber
 &~~\text{for}~~N_e=4,~M=3(N_e-1)=9.
 \end{align}
As shown above, this diagram is generated by removing two lines connecting particles $1\rightarrow3$ and $2\rightarrow4$ and replacing them by adding two lines connecting  $1\rightarrow2$ and $3\rightarrow4$. In algebraic form this can be expressed as  (note that division does not pose an issue as long as the resulting expression remains a polynomial),
 \begin{align}
|\psi\rangle=\frac{(a^\dagger_1-a^\dagger_2)(a^\dagger_3-a^\dagger_4)}{(a^\dagger_1-a^\dagger_3)(a^\dagger_2-a^\dagger_4)} \prod_{i<j}(a^\dagger_i-a^\dagger_j)^3|0\rangle.
 \end{align}
 Perhaps, it is more intuitive to write this as a LLL wave-function in real space,
  \begin{align}
\psi(z)=\frac{(z_1-z_2)(z_3-z_4)}{(z_1-z_3)(z_2-z_4)} \prod_{i<j}(z_i-z_j)^3 e^{-\sum_i |z_i|^2/4 l_B^2}.
 \end{align}
 The multiplicative factor $\frac{(z_1-z_2)(z_3-z_4)}{(z_1-z_3)(z_2-z_4)} $ can now be identified as a cross-ratio (invariant of conformal transformations). This suggestive form can be pushed further; As it is evident from diagrams, the Laughlin wave function multiplied by an \textit{arbitrary} polynomial function of cross-ratios corresponds to a ground state wave-function (under the condition that the final result remains a polynomial, i.e. number of lines connecting two vertices should be a non-negative integer),
\begin{align}
\psi(z)=P\Big(\frac{(z_i-z_j)(z_k-z_l)}{(z_i-z_k)(z_j-z_l)}\Big) \prod_{i<j}(z_i-z_j)^3 e^{-\sum_i |z_i|^2/4 l_B^2}.
 \end{align}
 In fact \text{all} ground state diagrams (wave functions) can be written in this form. 
 
 Note that ground state wave-functions satisfy the same properties as chiral conformal blocks, that is,
 \begin{align}
&P_\text{h}(\lambda a^\dagger)=\lambda^n P_\text{h}(a^\dagger),\newline \\ \nonumber
&P_\text{h}( a^\dagger+c)=P_h(a^\dagger).
 \end{align}

  This result then easily generalizes for any number of particles $N_e$ at any filling $m=\frac{M}{(N_e-1)}$, that is, the most general ground state wave-function of linear pseudo potential model is given by,
 \begin{align}\label{cfw}
\psi(z)=f\Big(\frac{(z_i-z_j)(z_k-z_l)}{(z_i-z_k)(z_j-z_l)}\Big) \prod_{i<j}(z_i-z_j)^{m} e^{-\sum_i |z_i|^2/4 l_B^2},
 \end{align}
Where $f\Big(\frac{(z_i-z_j)(z_k-z_l)}{(z_i-z_k)(z_j-z_l)}\Big)$ is an arbitrary function (not polynomial anymore) of cross ratios. We emphasize again that the total result is restricted to be a polynomial multiplied by the exponential factor (a LLL wave-function). Keeping this in mind it is clear that even (non-integer) rational values of $m$ do not pose any problem. The ground state energy is given by,
\begin{align}\label{en}
E_0(N_e,m)=- \frac{m N_e(N_e-1) }{2}.
\end{align}

Connection to chiral conformal blocks is now manifest; Apart from the trivial exponential factor, Eq.\eqref{cfw} corresponds to the most general homogeneous \text{polynomial} of degree $- E_0(N_e,m)$ that can be written as a correlation function of identical primary fields $\phi$ (with equal scaling dimensions) of a CFT (an arbitrary field theory invariant with respect to \textit{global} conformal transformations $SL(2,\mathbb{C})$)\cite{yellow},
\begin{align} 
\langle \phi(z_1 ) \phi(z_2) ... \phi(z_{N_e} )\rangle_{\text{CFT}}.
\end{align} 
For bosons/fermions, the symmetry/antisymmetry condition has to be externally imposed. This simple yet powerful and general result (in disk geometry) seems to have not been noticed before.

Before closing this section, let's consider an explicit example of a filling fraction where a polynomial Laughlin wave-function doesn't exist. Consider four particles $N_e=4$ at a filling $M=2(N_e-1)-1$. There are several diagrams associated with the ground state, for example,
\begin{align}
&\vcenter{\hbox{\includegraphics[width=0.7\columnwidth,keepaspectratio]{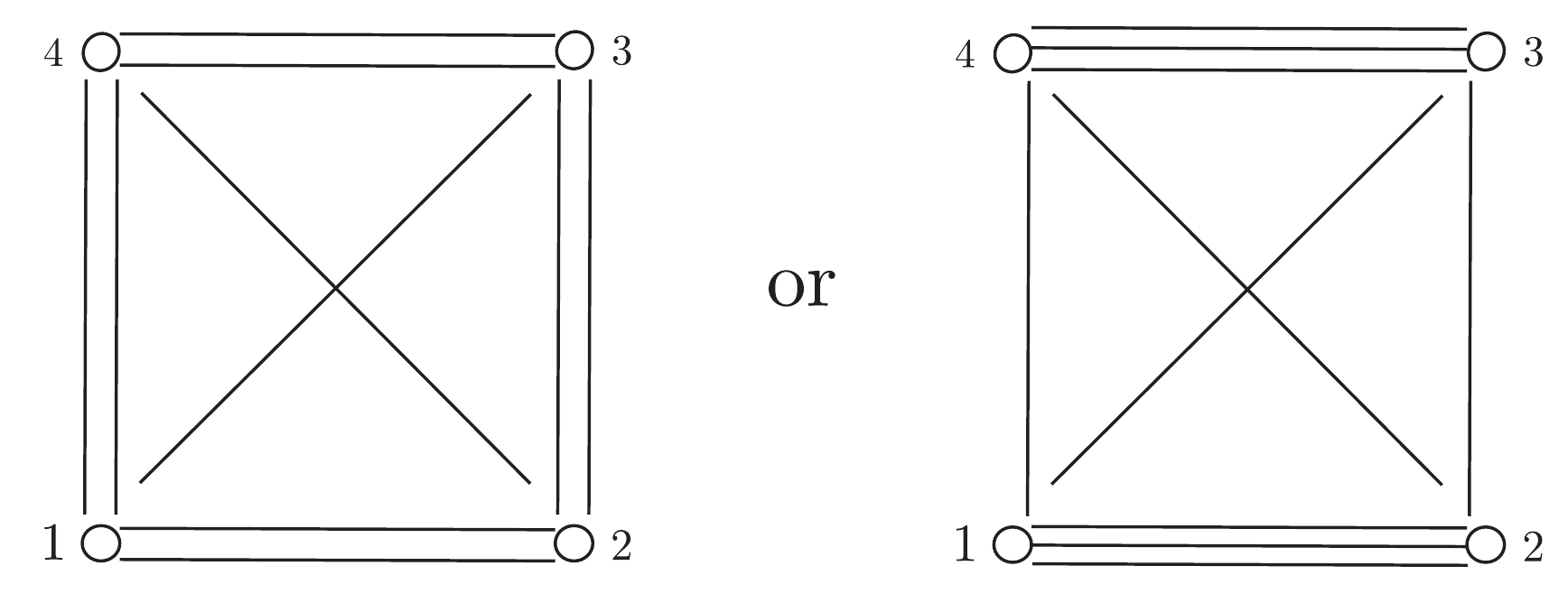}}}\\ \newline \nonumber
&~~~~~~\text{for}~~N_e=4,~M=2(N_e-1)-1=5.
\end{align}
However, in this case, if one considers only fermionic particles (anti-symmetric wave-functions) all ground state diagrams turn out to correspond to the same state, and so, the ground state is unique. Interestingly, this \text{unique} fermionic state is given by the half-filled Pfaffian wave function\cite{mooreread},
\begin{align}
&\vcenter{\hbox{\includegraphics[width=0.4\columnwidth,keepaspectratio]{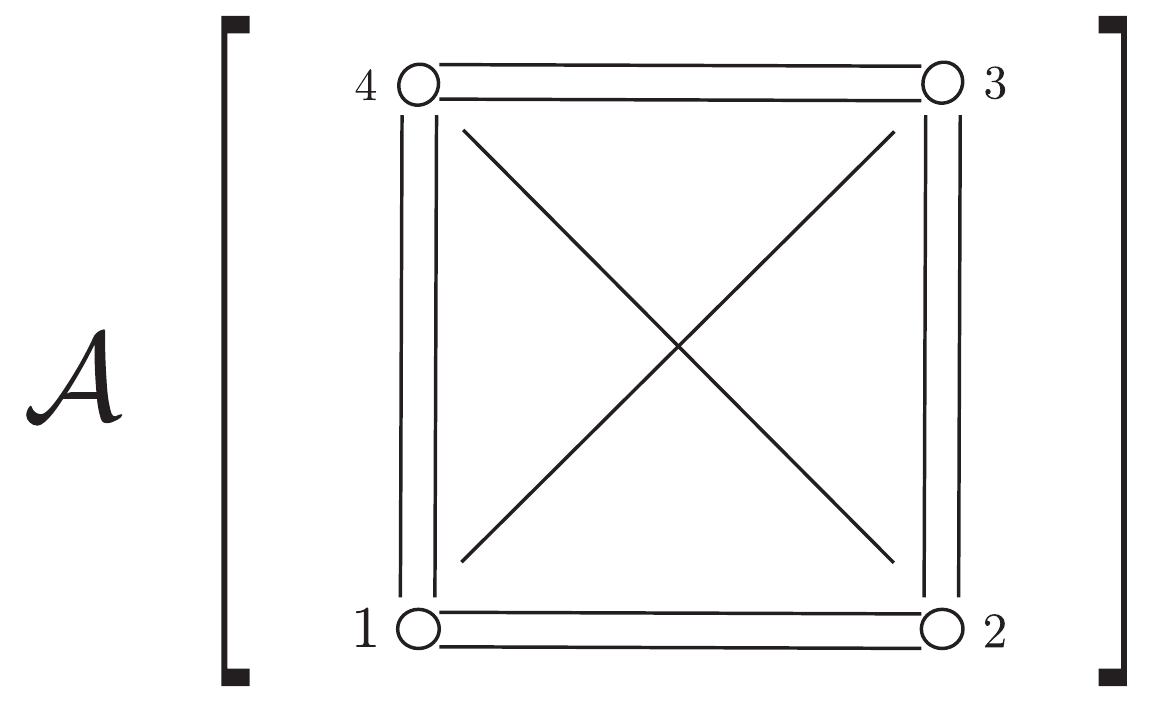}}} ~~\text{for}~~N_e=4,~M=2(N_e-1)-1=5
\\ \newline \nonumber&~~~~~= \text{Pf }(\frac{1}{z_i-z_j})  \prod_{i<j=1}^4(z_i-z_j)^2 e^{-\sum_i |z_i|^2/4 l_B^2}.
\end{align}
 Where the notation $\mathcal{A}$ means anti-symmetric over particle indices. That is for four \textit{fermions} $N_e=4$ at system size $M=2(N_e-1)-1=5$, the unique ground state wave-function is \text{exactly} given by the Moore-Read Pfaffian state (Note that this wave-function can also be written as Eq.\eqref{cfw}, with $m=5/3$). However, this results does not hold for larger number of particles (there the ground state becomes degenerate).

 \subsection{Quadratic Perturbation}\label{psd2}
 In the last section, we showed that the ground state subspace of the linear pseudopotential model coincides with the space of polynomial chiral conformal blocks. This space (particularly for a larger number of particles) is hugely degenerate. This is an artifact of the highly symmetric model we used. To remove this degeneracy, we add a small quadratic perturbation to the pseudopotential (Eq.\eqref{lin}),
 \begin{align}\label{lin2}
 - \frac{2}{N_e}N + \frac{4\alpha}{N_e^2}N^2.
\end{align} 
 The motivation for choosing this particular form of perturbation is to model a physically reasonable, \text{convex}, monotonically decreasing interaction potential, that is still analytically tractable.
 
 To first order in perturbation theory we need to diagonalize the quadratic perturbation $\frac{4\alpha}{N_e^2}N^2$ in the ground state subspace of the linear model. To this end, we need to consider matrix elements of the form,
 \begin{align}
\sum_{i<j} \frac{4\alpha}{N_e^2}  \langle l|  N_{ij}^2 |n \rangle,
 \end{align}
where $|n \rangle,|l \rangle$ are in the degenerate subspace. We can expand the expression above as (using translation invariance of conformal blocks $\sum_i a_i |n\rangle=0$),
 \begin{align}
&\sum_{i<j} \frac{4\alpha}{N_e^2}  \langle l|  N_{ij}^2 |n \rangle = \sum_{i<j} \frac{4\alpha}{N_e^2}  \langle l|  ((a^\dagger_i-a^\dagger_j)(a_i-a_j))^2  |n \rangle \\ \newline \nonumber
=&\langle l| \frac{4\alpha (N_e+1)}{N_e^2}     \sum_{i}(a_i^{\dagger} a_i)^{2} + \frac{4\alpha}{N_e^2} (\sum_i  a_i^{\dagger2} ) (\sum_i  a_i^{2} ) |n \rangle  \\ \newline \nonumber
&+ C(E_0(N_e,m)) \langle l| n \rangle
 \end{align}
 Where $C(E_0(N_e,m))$ is an inconsequential constant that is fixed by the ground state energy of the linear model. The first two terms above therefore define an effective Hamiltonian in the degenerate subspace,
 \begin{align}\label{efh}
 \frac{1}{4\alpha}H_{\text{ef}}=&  \frac{1}{N_e^2} (\sum_i  a_i^{\dagger2} ) (\sum_i  a_i^{2} ) + \frac{ (N_e+1)}{N_e^2}     \sum_{i}(a_i^{\dagger} a_i)^{2} \\ \nonumber \newline
 =& \mathcal{H}_1 + \mathcal{H}_2.
 \end{align}
 Where $\mathcal{H}_1$ and $\mathcal{H}_2$ correspond to the first and second term of the first line respectively. The conformal block wave-functions typically considered as trial states for quantum Hall states are obtained from the Moore-Read construction\cite{mooreread,read09,dubail}. These states are constructed as conformal blocks of chiral $U(1)\otimes \Psi$ (charge and statistics sectors) CFTs,
 \begin{align}
\psi_{\text{MR}}(z)=\frac{1}{Z_B} \langle \Big[\prod_{i} e^{i \sqrt{m} \varphi(z_i)} \times \Psi(z_i)\Big] e^{\frac{-i}{2\pi l_B^2} \int_\gamma d\tilde{z} \varphi (\tilde{z})}\rangle_\text{CFT},
 \end{align}
where the region $\gamma$ is a big enough disc to include all points $z_i$. $Z_B$ is a numerical factor that makes the bare trial wave function $\psi_{\text{MR}}$ normalized. $\varphi(z)$ is a free chiral boson, and $\Psi(z)$ is an abelian vertex operator in the (rational and unitary) statistics sector $\text{CFT}_\Psi$. Different choices of $\Psi$ correnpond to different trial wave-functions. Examples of which include Pfaffian\cite{mooreread} and Read-Rezayi\cite{rr} states.

 We show that in thermodynamic limit $N_e\rightarrow \infty$ these wave functions are \textit{exact eigenstates} of the Hamiltonian above (\textit{exact eigenstates} of both $\mathcal{H}_1$ and $\mathcal{H}_2$). Moreover, we show that these states are \textit{exact} ground states of $\mathcal{H}_1$. Note that $\mathcal{H}_2$ is an entirely single-body Hamiltonian $\mathcal{H}_2= \frac{N_e+1}{N_e^2}L_z^2\approx\frac{L_z^2}{N_e}$ (all two-body terms are included in $\mathcal{H}_1$). Hence, $\mathcal{H}_2$'s effect can be entirely subtracted by an external quadratic potential $-\frac{L_z^2}{N_e}$. Therefore, we can construct a Hamiltonian comprised entirely from two-body interactions and single-body potentials that has Moore-Read type conformal blocks as its exact ground state in the thermodynamic limit. The possibility of residual degeneracies can be ruled out here. This is the central result of this work.

To calculate the average of the first term it is useful to write it in real space form as,
\begin{align}\label{e1}
 &\frac{1}{N_e^2}   \langle (\sum_i  a_i^{\dagger2} ) (\sum_i  a_i^{2} ) \rangle =  \\ \newline \nonumber & \frac{1}{N_e^2} \langle \Big [ (\sum_i  a_i^{2} ) (\sum_i  a_i^{\dagger2} ) - \sum_i (4 a^\dagger_i a_i + 2 ) \Big]\rangle= \\ \newline \nonumber
 &\frac{1}{N_e^2 Z_B}\int \prod_i^{N_e} dz_i \psi^*_{\text{MR}}(z) (\sum_i z^{*2}_i) (\sum_i z^{2}_i) \psi_{\text{MR}}(z) \\ \newline \nonumber 
 &- \frac{1}{N_e^2}(2 m N_e (N_e-1) + 2 N_e)
\end{align}
 Inner products of the type above have been extensively studied; Originally Ref.~\onlinecite{wen} used plasma analogy arguments to calculate inner products of this form for the Laughlin wave function\cite{wen}. Later on, this results were extended and expanded to more general (non-abelian) Moore-Read conformal blocks obeying a ``generalized screening condition" using CFT techniques in Ref. ~\onlinecite{dubail}. These results were verified and studied in more detail in Refs.~\onlinecite{simon1,simon2,jackson}. The relevant part of those works for us is as follows; defining ``collective" coordinates  $S_k=\sqrt{m}\sum_i\frac{z^k_i}{(m N_e)^{k/2}}$ we have,
\begin{align}\label{read}
&\lim_{N_e\rightarrow \infty}\frac{1}{Z_B}\int \prod_i^{N_e} dz_i \psi^*_{\text{MR}}(z) S_{k_1}^* ... S_{k_p}^* S_{k'_1} ... S_{k'_q}  \psi_{\text{MR}}(z)   \\ \newline \nonumber 
& =\langle 0 |J_{k'_1} ... J_{k'_q} J_{-k_1} ... J_{-k_p}  | 0 \rangle_{\text{CFT}} + \mathcal{O}(\frac{1}{\sqrt{N}})
\end{align}
Where $J_n$ operators are modes of a $U(1)$ current in the CFT satisfying,
\begin{align}
[J_n,J_m]=n ~\delta_{n+m,0}.
\end{align}
The vacuum state  $|0\rangle$ is annihilated by all positive current modes $J_n$ ($n>0$). 

Using Eq.\eqref{read} we can now explicitly calculate Eq.\eqref{e1} (for Moore-Read conformal blocks) in the thermodynamic limit,
\begin{align}\label{imp}
&\lim_{N_e\rightarrow\infty} \langle  \psi_{\text{MR}} |  \mathcal{H}_1|  \psi_{\text{MR}}  \rangle =\newline \\ \nonumber
 &\lim_{N_e\rightarrow\infty} \frac{1}{N_e^2}  \langle  \psi_{\text{MR}} | (\sum_i  a_i^{\dagger2} ) (\sum_i  a_i^{2} )| \psi_{\text{MR}} \rangle= 0 + \mathcal{O}(\frac{1}{\sqrt{N_e}}).
\end{align}
Which means this class of wave-functions are the \textit{exact ground states} of $\mathcal{H}_1$ in the thermodynamic limit (since $\mathcal{H}_1$ is positive). Note that for a generic state, this expectation value is expected to be of the order of $\mathcal{O} (N_e^2)$.

We now proceed to discussing $\mathcal{H}_2$. It is useful to rewrite $\langle \mathcal{H}_2 \rangle$ as (in the thermodynamic limit),
\begin{align}
\lim_{N_e\rightarrow\infty}\langle \mathcal{H}_2 \rangle= \frac{1}{N_e}  \langle\sum_{i}(a_i^{\dagger} a_i)^{2} \rangle= \frac{1}{9 N_e} \langle[\sum_i a_i^3 , \sum_j a_j^{\dagger3}]\rangle.
\end{align}
Using  Eq.\eqref{read} for Moore-Read conformal blocks, this can be further simplified to,
\begin{align}
&\lim_{N_e\rightarrow\infty} \langle  \psi_{\text{MR}} |  \mathcal{H}_2|  \psi_{\text{MR}}  \rangle=  \frac{1}{9 N_e}  \langle  \psi_{\text{MR}} | [\sum_i a_i^3 , \sum_j a_j^{\dagger3}]|  \psi_{\text{MR}}  \rangle \\ \newline \nonumber
&=\frac{1}{3} m^2 N_e^2.
\end{align}
 To get a sense of scale, it is helpful to note in the conformal subspace $\langle \mathcal{H}_2 \rangle>\frac{1}{4} m^2 N_e^2$ is bounded (since $\langle a_i^\dagger a_i \rangle = \frac{m}{2} N_e$).

To see that Moore-Read conformal blocks are exact eigenstates note that (derivation similar to the previous equation),
\begin{align}
&\lim_{N_e\rightarrow\infty}  \frac{\langle  \psi_{\text{MR}} |  \mathcal{H}_2^2 |  \psi_{\text{MR}}  \rangle }{ \langle  \psi_{\text{MR}} |  \mathcal{H}_2 |  \psi_{\text{MR}}  \rangle^2}   =1.
\end{align}
Thus, in the thermodynamic limit Moore-Read conformal blocks become exact eigenstates of $\mathcal{H}_2$. As mentioned earlier, $\mathcal{H}_2$ is an entirely single particle term, and therefore, it's coefficient can be manipulated by an external $L_z^2$ potential. 

Therefore, in the thermodynamic limit, the class of Hamiltonians (including only two-body interactions and external single-body potentials),
\begin{align}
H=& - \sum_{j<k}\frac{2}{N_e}N_{jk} +  \sum_{j<k} \frac{4\alpha}{N_e^2}N_{jk}^2 + W \sum_{j} \Theta(a^\dagger_j a_j-M) \\ \newline \nonumber
 &- \sum_j \frac{\beta}{N_e} N_j^2,
\end{align} 
 have Moore-Read conformal blocks (e.g. Laughlin, Pfaffian and Read-Rezayi states) as their exact eigenstates in the thermodynamic limit. Moreover, by tuning the external single-body potential $L_z^2$, that is setting  $\beta=1$, we can enforce Moore-Read conformal blocks to become exact ground states in the thermodynamic limit. However, the possibility of residual degeneracies can not be ruled out here.  

\section{Discussion and Conclusion}\label{disc}
In conclusion, we have introduced a  model of particles in a LL interacting with long-rage almost linear pseudo-potential interactions as well as an externally imposed potential that has Moore -Read type conformal blocks (e.g. Laughlin, Pfaffian, and Read-Rezayi states) as its exact ground state in the thermodynamic limit. We have identified a connection between the ground-state subspace of the linear pseudo-potential model with the space of polynomial conformal blocks of CFTs. As opposed to earlier models for FQH effect, our model does not have explicit filling dependence, that is, the same model can be used for all filling fractions. One can then hope to study FQH transitions within our model

A number of important unanswered questions remain, in particular; We have not ruled out the possibility of residual ground state degeneracies in the thermodynamic limit. We have not addressed the nature and finiteness of the spectral gap in the thermodynamic limit (there are hints that this issue might be related to unitarity of the statistics sector $\text{CFT}_\Psi$\cite{read8}). We have not discussed the existence (or lack thereof) GMP like magneto-roton modes\cite{gmp} within our  model.

Our results shine new light on the emergence of conformal block wave-functions in FQH systems from a bulk perspective. We hope that our work can stimulate further work studying more general (and perhaps realistic) pseudopotential models using connections with CFT and perturbation theory.

 \section*{Acknowledgement}
I am grateful to Ali Lavasani, Jay Sau, Maissam Barkeshli, Jerome Dubail, and Steve Simon for valuable comments on the manuscript. This work was supported by JQI-NSF-PFC and the National Science Foundation NSF DMR-1555135.

\bibliography{library}

\begin{widetext}

\section{Supplementary Material: Numerical Results for Small Systems}\label{num}

Here we provide numerical results obtained using exact diagonalization for small systems. We work in the ground state subspace of the linear model (space of conformal blocks), and diagonalize $\mathcal{H}_1$ and $\mathcal{H}_2$ within that subspace. Note that we do not impose symmetry/anti-symmetry constraints on our wave-functions (this can be relevant, for example, for spinful FQHE of electrons). Moreover, as opposed to most FQHE numerics we work in disk geometry, for this reasons, we deal with an unusually large Hilbert-space and are therefore limited to quite small systems of $N=4 ~\text{and} ~5$. Sample results are plotted in Figs.\ref{fig:E1.pdf} and \ref{fig:E2.pdf}. 

As shown in Fig.\ref{fig:E1.pdf}, and in agreement with our analytical results, the Moore-Read conformal blocks seem to minimize $\mathcal{H}_1$ even for quite small system sizes. We have considered three cases; 1. $N_e=5$ particles at full filling $m=1$, in this case only one anti-symmetric (fermionic) wave-function exists, however since we are not demanding anti-symmetric/symmetric wave-functions the Hilbert space is larger (ground state subspace is $16$ dimensional in this case). In this case the Laughlin $m=1$ wave-function (Vandermonde Polynomial) exactly minimizes $\mathcal{H}_1$ even for finite sizes. 2. $N_e=4$ particles at half-filling $m=2$, in this case the bosonic $m=2$ Laughlin state seems to be already a good approximation for the ground state of $\mathcal{H}_1$. Moreover, it seems to be separated from the rest of the states by a gap. 3. $N_e=4$ particles at $M=5$, as mentioned earlier, in this case only one fermionic state (Pfaffian) exists (without demanding symmetry the space is larger). Again, Pfaffian state seems to be a good approximation for the ground state of $\mathcal{H}_1$.

We also looked at the spectrum of $\mathcal{H}_2$ in simple cases (Fig.\ref{fig:E2.pdf}). We have considered the same three cases as above. $N_e=5$ particles at full filling $m=1$ case is worth special attention, in this case the Laughlin $m=1$ wave-function (Vandermonde Polynomial) exactly maximizes $\mathcal{H}_1$. Note that for the Laughlin $m=1$ state, finite size corrections of Eq.\eqref{read} vanish. This is suggestive that the $m=1$ finite size results can be extended to all  Moore-Read conformal blocks at thermodynamic limit. For this reason, we conjecture that Moore-Read conformal blocks maximize $\mathcal{H}_2$ in the thermodynamic limit. We attribute the deviations from this result in other filling fraction to finite size effects (finite size effects are expected to be stronger for Pfaffian state compared with the Laughlin state\cite{dubail}). However, we emphasize that the validity of this conjecture does not alter the results of this work.
 \begin{figure}[ht]
\centering
\includegraphics[width=\columnwidth,keepaspectratio]{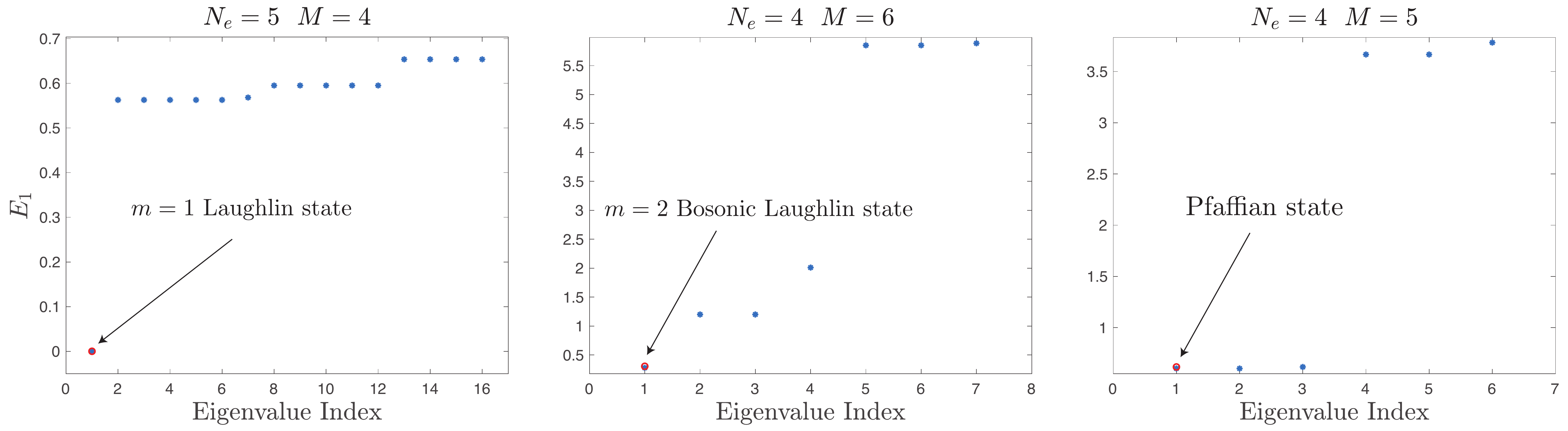} 
\caption{Spectrum of $\mathcal{H}_1$ in the ground state subspace of the linear model. Red circles mark $\langle\mathcal{H}_1\rangle_\text{MR}$ for Moore-Read conformal blocks. Note that we are not restricted to symmetric/antisymmetric states. \label{fig:E1.pdf}}
\end{figure}

 \begin{figure}[h]
\centering
\includegraphics[width=\columnwidth,keepaspectratio]{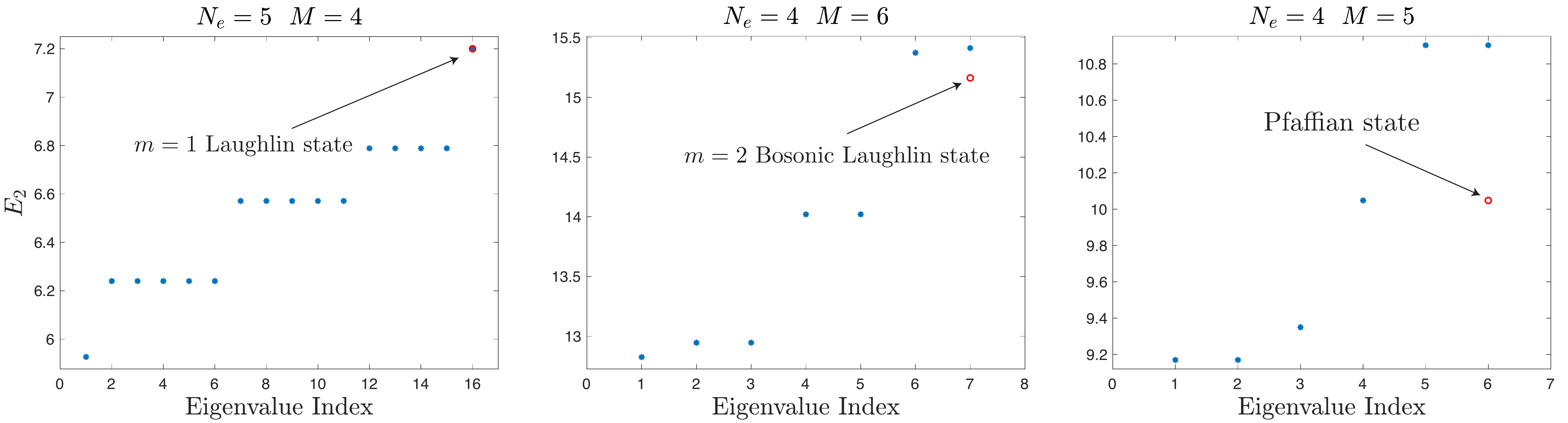} 
\caption{Spectrum of $\mathcal{H}_2$ in the ground state subspace of the linear model. Red circles mark $\langle\mathcal{H}_2\rangle_\text{MR}$ for Moore-Read conformal blocks. Note that we are not restricted to symmetric/antisymmetric states. \label{fig:E2.pdf}}
\end{figure}
\end{widetext}
\end{document}